\documentclass[aip,graphicx,reprint,longbibliography,noeprint]{revtex4-1}
\usepackage[pdftex]{graphics}

\usepackage{amssymb,amsmath}
\usepackage{bm}
\usepackage{siunitx}

\newcommand{\rd}{\mathrm{d}}
\newcommand{\ii}{\mathrm{i}}
\newcommand{\e}{\mathrm{e}}
\newcommand{\Tr}{\mathrm{Tr}}
\renewcommand{\Re}{\operatorname{Re}}
\renewcommand{\Im}{\operatorname{Im}}

\usepackage{dcolumn} 
\newcolumntype{d}[1]{D{.}{.}{#1}} 

\begin{document}

\title{Exciton dynamics from the mapping approach to surface hopping: Comparison with Förster and Redfield theories}
\author{Johan E. Runeson}
\email{johan.runeson@chem.ox.ac.uk}
\affiliation{Department of Chemistry, University of Oxford, Physical and Theoretical Chemistry Laboratory, South Parks Road, Oxford, OX1 3QZ, UK}
\author{Thomas P. Fay}
\affiliation{Department of Chemistry, University of California, Berkeley, California 94720, USA\looseness=-1}
\author{David E. Manolopoulos}
\affiliation{Department of Chemistry, University of Oxford, Physical and Theoretical Chemistry Laboratory, South Parks Road, Oxford, OX1 3QZ, UK}

\begin{abstract}
We compare the recently introduced multi-state mapping approach to surface hopping (MASH) with the F\"orster and Redfield theories of excitation energy transfer. Whereas F\"orster theory relies on weak coupling between chromophores, and Redfield theory assumes the electronic excitations to be weakly coupled to fast chromophore vibrations, MASH is free from any perturbative or Markovian approximations. We illustrate this with an example application to the rate of energy transfer in a Frenkel-exciton dimer, showing that MASH interpolates correctly between the opposing regimes in which the F\"orster and Redfield results are reliable. We then compare the three methods for a realistic model of the Fenna--Matthews--Olson complex with a structured vibrational spectral density and static disorder in the excitation energies. In this case there are no exact results for comparison so we use MASH to assess the validity of F\"orster and Redfield theories. We find that F\"orster theory is the more accurate of the two on the picosecond timescale, as has been shown previously for a simpler model of this particular light-harvesting complex. We also explore various ways to sample the initial electronic state in MASH and find that they all give very similar results for exciton dynamics.
\end{abstract}

\maketitle

\section{Introduction}
Revealing the mechanisms of excitation energy transfer is fundamental to the study of biological photosynthetic systems and has motivated methodological development for several decades.\cite{Mirkovic2017review,segatta2019review}
Excitonic systems typically consist of an assembly of chromophores (sites) embedded in a protein or solvent environment. Traditionally, excitation energy transfer has been divided into two regimes, depending on the relative magnitudes of the vibrational relaxation time and the inverse of the inter-site coupling strength.
\cite{may_kuehn,ishizaki2010PCCP} For chromophores that are weakly coupled to each other (but have rapid vibrational relaxation, which usually means strong coupling to the environment) the transfer is referred to as `incoherent' and proceeds via hopping between sites. This regime can be well described by F\"{o}rster theory. In the opposite limit of slow vibrational relaxation (or strong coupling between the chromophores), the transfer is referred to as `coherent' and proceeds between delocalized exciton states. This regime can be described by a perturbative master equation such as Redfield theory.
We emphasize that in either regime, the transfer of population is often well described by a kinetic rate equation, and that the electronic coherences are both small and short-lived in an appropriate electronic basis.\cite{faraday2020coherence,cao2020review}

The main challenge is that many systems of interest tend to appear in the intermediate regime where the two timescales are comparable.\cite{Scholes2018jpcl,Sneyd2022exciton}
For simple system--bath models one can solve the dynamics non-perturbatively using fully quantum methods such as the hierarchical equations of motion (HEOM).\cite{tanimura1989heom} Such a study of a two-site model has shown that exciton transfer is fastest in the intermediate regime, where the rate is not adequately captured by either of the perturbative theories mentioned above.\cite{ishizaki2009unified} However, fully quantum methods like HEOM and quasi-adiabatic path integrals\cite{Makri1995a,Makri2020smatpi} are only practical for harmonic models and they can be hard to converge for strong system--environment couplings. It is therefore important to develop accurate methods that can describe nonadiabatic transitions more generally, including in systems with anharmonic atomistic potentials.

So far, the most successful strategy to model nonadiabatic transitions has been Tully's fewest switches surface hopping (FSSH).\cite{tully1990hopping} Since it was first proposed in 1990, this stochastic algorithm has become immensely popular in photochemistry and is widely implemented in open software. However, it continues to suffer from long-standing issues often referred to as `overcoherence', despite much effort on the development of more or less {\em ad hoc} `decoherence corrections'.\cite{wang2020review}

Recently, Mannouch and Richardson have proposed a different strategy based on a phase-space mapping of two-state systems onto a spin degree of freedom.\cite{Mannouch2023mash} This so-called `mapping approach to surface hopping' (MASH) uses a deterministic algorithm that, in contrast to FSSH, hops to the adiabatic surface with the largest instantaneous population. This strategy has many appealing features. Firstly, it removes all ambiguity about the need for velocity rescaling/reversal for successful/unsuccessful hops. Secondly, it replaces {\em ad hoc} decoherence corrections with a rigorous `quantum jump' procedure, even without which it has been found to be more accurate than FSSH in a range of benchmark applications.\cite{Mannouch2023mash} Thirdly, unlike FSSH, it can correctly describe the transition between adiabatic and non-adiabatic rates in the spin-boson model and it recovers Marcus theory in the limit of a perturbative inter-state coupling.\cite{Lawrence2023mash}
Finally, MASH has been proven to relax to the correct quantum--classical equilibrium distribution for ergodic systems, a feature that is not shared by any other nonadiabatic trajectory method.\cite{Amati2023thermalization}

In its original formulation, MASH was limited to systems with two electronic states, but an adaptation to multiple states has recently been proposed.\cite{Runeson2023mash} Compared to the original approach, the multi-state formulation differs in the way it calculates electronic observables. The multi-state estimators are constructed to be equivariant under unitary basis transformations, meaning that populations and coherences are treated on the same footing and observables can be directly evaluated in any basis. With these estimators, multi-state MASH provably relaxes to the correct quantum--classical equilibrium for a general $N$-state system, in any basis.\cite{Runeson2023mash} 

Unfortunately, the equivariant multi-state formulation of MASH\cite{Runeson2023mash} does not reduce to the original formulation\cite{Mannouch2023mash} in the two-state case. The two methods have nevertheless been shown to be of comparable accuracy for a wide variety of two-state systems.\cite{Runeson2023mash} A noteworthy exception was found for a spin-boson model in the Marcus inverted regime, for which the original formulation was more accurate than the multi-state formulation.\cite{Runeson2023mash} 
However, we shall show below that this conclusion does not hold generally, and that for a two-site exciton transfer model the two formulations lead to essentially the same rates, even in the inverted regime.

In this article, we apply multi-state MASH to a more challenging set of exciton systems than previously considered. In particular, we investigate the transition between the regimes of `coherent' and `incoherent' rate theories and compare with fully quantum benchmark results. We then go on to investigate exciton transfer in the Fenna--Matthews--Olson (FMO) complex, using an experimentally measured spectral density to account for coupling to vibrations, and including static disorder in the site energies. Comparison to simple rate theories indicates that FMO is better described by F\"{o}rster theory than by Redfield theory, despite the popularity of the latter in the exciton literature.

We also analyse in detail the choice of initial conditions in multi-state MASH. The choice used in Ref.~\onlinecite{Runeson2023mash} is not unique and, more importantly, not basis-equivariant, in contrast to the treatment of electronic observables. We exemplify why this could become a problem for a system with three states, and consider alternative sets of initial conditions. In particular, we present an approach that overcomes the objection for three states and restores the equivariance of the initial distribution. Upon comparison for exciton transfer in the dimer and FMO models, however, we find that the initial distribution has little practical influence on the dynamics. Based on these results, we conclude that using the simplest set of initial conditions is well-justified for the kind of system we consider here.

\section{Perturbative rate theories}
As a prototypical model for exciton energy transfer, we consider the Frenkel-exciton Hamiltonian,
\begin{equation}\label{eq:FEham}
    H = H_{\rm S} + H_{\rm B} + H_{\rm SB}.
\end{equation}
The first term is the system Hamiltonian in the basis of localized pigment (`site') excitations
\begin{equation}
    H_{\rm S} = \sum_{n=1}^N \epsilon_n|n\rangle\langle n| + \sum_{n>m}J_{nm}(|n\rangle\langle m|+|m\rangle\langle n|),
\end{equation}
where $\{\epsilon_n\}$ and $\{J_{nm}\}$ are the site energies and the inter-site couplings, respectively. To model interaction with vibrational and solvent degrees of freedom, the on-diagonal site energies are linearly coupled to a harmonic bath,
\begin{equation}
    H_{\rm B} = \sum_{n=1}^N \sum_{j=1}^f \left(\frac{p_{j,n}^2}{2} + \frac{1}{2}\omega_j^2 q_{j,n}^2 \right),
\end{equation}
\begin{equation}
    H_{\rm SB} = \sum_{j=1}^f \sum_{n=1}^N \kappa_j q_{j,n} |n\rangle\langle n|.
\end{equation}
Here, $p_{j,n}$ and $q_{j,n}$ are mass-scaled ($m_{j,n}=1$) momentum and coordinate variables for the vibrational modes.
The bath frequencies and vibrational couplings are specified through the spectral density
\begin{equation}
    J(\omega) = \frac{\pi}{2} \sum_j \frac{\kappa_j^2}{\omega_j}\delta(\omega-\omega_j),
\end{equation}
and all sites are assumed to be coupled to identical and independent baths. 
As a measure of the overall system--bath coupling strength, we define the bath reorganization energy 
\begin{equation}
    \lambda = \frac{1}{\pi}\int_0^\infty \frac{J(\omega)}{\omega}\rd \omega = \sum_j \frac{\kappa_j^2}{2\omega_j^2}.
\end{equation}
The system--bath interaction involves the bath operator $B_n=\sum_j \kappa_j q_{j,n}$, which is the energy gap between a localized excitation on site $n$ and the ground state. In the following, we drop the index $n$ since the baths are identical. To characterize the dynamics of the bath, it is useful to define the 
autocorrelation function of the bath operator,
\begin{equation}
    C(t) = \Tr_{\rm B}[\e^{-\beta {H}_{\rm B}}B(0)B(t)],
\end{equation}
where $\beta=1/(k_{\rm B}T)$ and the time-dependence refers to dynamics under ${H}_{\rm B}$. Using known expressions for the thermal correlation functions of a harmonic oscillator, one can show that
\begin{align}\label{eq:Ct}
    C(t) &= \frac{1}{\pi}\int_0^\infty \rd\omega \, J(\omega)\left[\coth\left(\frac{\beta\omega}{2}\right) \cos\omega t + \ii \sin\omega t \right]    \\
     &= \frac{1}{\pi}\int_0^\infty \rd\omega \, J(\omega)\left[(1+n(\omega))\e^{\ii \omega t} + n(\omega)\e^{-\ii\omega t} \right]    
\end{align}
where $n(\omega)=1/(\e^{\beta\omega}-1)$ is the Bose--Einstein distribution and throughout we use units where $\hbar=1$.



\subsection{Redfield theory}\label{sec:redfield}
In the limit of weak system--bath coupling, the effect of the bath on the excitonic system can be described by a second-order perturbative master equation. The perturbation expansion is usually truncated in the eigenbasis of $H_{\rm S}$, which is called the \emph{exciton} basis. Diagonalizing the system Hamiltonian gives $H_{\rm S} = \sum_\mu \omega_\mu|\mu\rangle\langle \mu|$, where $\omega_\mu$ is an eigenenergy and $|\mu\rangle = \sum_n U^{-1}_{\mu n}|n\rangle = \sum_n U_{n\mu}|n\rangle$ is an exciton state. We also assume that $C(t)$ decays faster than the timescale of the system dynamics.
Under these two assumptions (a weak and fast bath), a standard textbook derivation\cite{may_kuehn} leads to a Markovian master equation for the reduced density matrix of the system,
\begin{equation}\label{eq:redfield}
    \frac{\partial}{\partial t}\rho_{\mu\nu}(t) = -\ii \omega_{\mu\nu}\rho_{\mu\nu}(t)+\sum_{\mu'\nu'}R_{\mu\nu\mu'\nu'}\rho_{\mu'\nu'}(t),
\end{equation}
where $\omega_{\mu\nu}=\omega_\mu-\omega_\nu$ is the energy gap between the corresponding eigenstates of $H_{\rm S}$. The second term involves the Redfield tensor
\begin{align}
    R_{\mu\nu\mu'\nu'} &= \Gamma_{\nu'\nu\mu\mu'} + \Gamma^*_{\mu'\mu\nu\nu'}\nonumber \\
    &-\delta_{\nu\nu'}\sum_\kappa\Gamma_{\mu\kappa\kappa\mu'} - \delta_{\mu\mu'}\sum_\kappa\Gamma^*_{\nu\kappa\kappa\nu'},
\end{align}
which is expressed in terms of the damping tensor
\begin{equation}
    \Gamma_{\mu\nu\mu'\nu'}=\sum_{n}\langle \mu|n\rangle\langle n|\nu\rangle\langle \nu'|n\rangle\langle n|\mu'\rangle \tilde{C}(\omega_{\mu\nu}).
\end{equation}
Here, $\tilde{C}(\omega)$ is the Fourier--Laplace transform of the bath correlation function,
\begin{equation}
    \tilde{C}(\omega) = \int_0^\infty \rd t \;\e^{-\ii\omega t} C(t).
\end{equation}
Inserting the expression for $C(t)$ from Eq.~\eqref{eq:Ct} gives
\begin{align}
    \Re \tilde{C}(\omega) &= J(\omega)(1+n(\omega)) + J(-\omega)n(-\omega) \label{eq:ReC} \\
    \Im \tilde{C}(\omega) &= \frac{1}{\pi}\mathrm{P}\int_{-\infty}^\infty \rd \omega'\; \frac{\Re \tilde{C}(\omega)}{\omega'-\omega},
\end{align}
where $\mathrm{P}$ denotes principal value and, with the notation used in this paper, $J(\omega<0)=0$, so only one term on the right-hand side of Eq.~\eqref{eq:ReC} is non-zero. 
Finally, if the diagonal elements of ${\rho}$ (the exciton populations) are only weakly influenced by the off-diagonal elements (the exciton coherences), one can replace Eq.~\eqref{eq:redfield} by a kinetic rate equation for the populations (the secular approximation),
\begin{equation}
    \frac{\partial}{\partial t}\rho_{\mu\mu}(t) = -\sum_{\mu \neq \nu} (k_{\nu\to\mu}\rho_{\nu\nu}-k_{\mu\to\nu}\rho_{\mu\mu})
\end{equation}
with the Redfield rate constants
\begin{equation}
    k_{\mu\to\nu}=2\sum_n\langle \mu|n\rangle\langle n|\nu\rangle\langle \nu|n\rangle\langle n| \mu\rangle \Re \tilde{C}(\omega_{\mu\nu}).
\end{equation}

\subsection{F\"{o}rster theory}
In the opposite limit of strong system--bath coupling, one may instead pick the perturbation parameter to be the intersite coupling $J_{nm}$. This is the F\"{o}rster-type incoherent hopping limit, in which the subsystem follows a kinetic rate equation in the \emph{site} basis. The F\"{o}rster rate constants are\cite{yang2002}
\begin{equation}\label{eq:forsterk1}
    k_{n\to m} = 2|J_{nm}|^2 \Re \int_0^\infty \rd t\; F_n^*(t)A_m(t)
\end{equation}
where $F_n(t)$ and $A_m(t)$ are the flourescence and absorption lineshape functions, which based on the cumulant expansion technique can be written as
\begin{align}
    F_n^*(t) &= \e^{+\ii(\epsilon_n-\lambda)t - g(t)} \\
    A_m(t) &= \e^{-\ii(\epsilon_m+\lambda)t - g(t)}.
\end{align}
The function $g(t)$ is
\begin{equation}
    g(t) = \int_0^t \rd t_1 \int_0^{t_1}\rd t_2\, C(t_2),
\end{equation}
where $C(t)$ is the bath correlation function in Eq.~\eqref{eq:Ct}.
For the Frenkel-exciton model with identical baths, the rate expression in Eq.~\eqref{eq:forsterk1} reduces to
\begin{equation}
    k_{n\to m} = 2|J_{nm}|^2 \Re \int_0^\infty \rd t\; \e^{\ii(\epsilon_n-\epsilon_m)t - 2g(t)},
\end{equation}
where
\begin{equation}
    g(t) = \sum_j \frac{\kappa_j^2}{2\omega_j^3}\left( \coth\left(\frac{\beta \omega_j}{2}\right)(1-\cos\omega_j t) + \ii \sin\omega_j t\right).
\end{equation}

\section{Multi-state MASH}\label{sec:mash}
An alternative strategy is to simulate the bath dynamics explicitly with surface hopping. To this end, we rewrite the Frenkel-exciton Hamiltonian in Eq.~\eqref{eq:FEham} as
\begin{equation}
    \hat{H}(p,q) = \sum_{j,n} \frac{p_{j,n}^2}{2} + \hat{V}(q),
\end{equation}
where
\begin{equation}
    \hat{V}(q) = \sum_{nm} V_{nm}(q)|n\rangle \langle m|
\end{equation}
is a (diabatic) potential operator with matrix elements
\begin{equation}
    V_{nm}(q) = \bigg(\epsilon_n+\sum_j \kappa_j q_{j,n} + \sum_{j,l}\frac{1}{2}\omega_j^2 q_{j,l}^2\bigg)\delta_{nm} + J_{nm}.
\end{equation}
Surface hopping is almost always run on adiabatic surfaces, i.e., in the eigenbasis of $\hat{V}(q)$,
\begin{equation}
    \hat{V}(q) = \sum_a V_a(q)|a(q)\rangle\langle(q)|.
\end{equation}
We will refer to $p, q$ as \emph{nuclear} variables (physically, they represent intramolecular vibrations as well as collective modes of the solvent). Their dynamics is coupled to the electronic wavefunction $|\psi\rangle$, which can be expanded in the diabatic or the adiabatic basis as
\begin{equation}
    |\psi\rangle = \sum_n c_n |n\rangle = \sum_a c_a |a(q)\rangle.
\end{equation}
The coefficients obey the Schr\"{o}dinger equation
\begin{equation}
    \dot{c}_n = -\ii \sum_m V_{nm}(q)c_m
\end{equation}
or (equivalently)
\begin{equation}
    \dot{c}_a = -\ii V_a(q)c_a - \sum_j \dot{q}_j \sum_b d_{ab}^j(q)c_b,
\end{equation}
where $d_{ab}^j(q)=\langle a(q)|\nabla_j|b(q)\rangle$ is a nonadiabatic coupling matrix element. (Here and in the rest of this section, the index $j$ runs over the nuclear degrees of freedom of all sites, not just one.)

In MASH, the nuclear trajectories evolve according to the classical equations of motion
\begin{subequations}
\begin{align}
    \dot{q}_j &= p_j/m_j \\
    \dot{p}_j &= -\langle a(q)|\nabla_j \hat{V}|a(q)\rangle
\end{align}
\end{subequations}
where $a$ is the adiabatic state with the largest instantaneous population $|c_a|^2$. Effectively, this means that the nuclei evolve on a potential with abrupt steps. By introducing the classical state projectors,
\begin{equation}
    \Theta_n = 
    \begin{cases} 
    1 & \text{if } |c_n|^2>|c_k|^2 \; \forall k\neq n \\
    0 & \text{otherwise}, 
    \end{cases}
\end{equation}
one can express the effective potential perceived by the nuclei as
\begin{equation}\label{eq:Veff}
    V_{\rm eff} = \sum_a V_a(q)\Theta_a(c).
\end{equation}
At each instant, precisely one of the $\Theta_a(c)$ factors in the sum of Eq.~\eqref{eq:Veff} is non-zero. Whenever a new state $b$ reaches a higher population than the current state $a$, the nuclei meet a potential step $V_b-V_a$. When crossing such a potential step, the momentum is rescaled so as to conserve energy, and if there is insufficient kinetic energy to overcome the step, the momentum is instead reversed. In multi-state MASH, the momentum component subject to rescaling/reversal is the projection onto the direction of a vector $v$ with elements\cite{Runeson2023mash} 
\begin{equation}
    v_j = \sum_{a'} \Re\left[ c^*_{a'}(d_{a'a}^jc_a - d_{a'b}^jc_b)\right].
\end{equation}
Because the hops occur deterministically, it is straightforward to adjust the timestep if necessary to better resolve a given hopping event (whereas in a stochastic algorithm, changing the timestep would change the locations of the hops).


\subsection{Estimators}
Having defined the dynamics of each trajectory, we next address how to measure observables. 
Consider a process starting in a pure electronic state $|i\rangle\langle i|$ with a (normalized) nuclear density $\rho_{\rm B}(p,q)$. The time-dependent expectation value of an observable $O$ is then
\begin{equation}
    \langle O(t)\rangle = \Tr[\rho_{\rm B}(p,q)|i\rangle \langle i|(0)\, \hat{O}(t)],
\end{equation}
where the trace runs over nuclear as well as electronic degrees of freedom. The corresponding expression in multi-state MASH is the phase-space integral
\begin{equation}\label{eq:Ot}
    \langle O(t)\rangle \approx \frac{\int \rd p\, \rd q \int_{|c|=1}\rd c \, \rho_{\rm B}(p,q) \rho_i(c)\,O(p_t,q_t,c_t)}{\int \rd p\, \rd q \int_{|c|=1}\rd c \, \rho_{\rm B}(p,q) \rho_i(c)}
\end{equation}
where $\int_{|c|=1}\rd c$ is an integral over all normalized electronic wavefunctions. A simple way to sample this integral is generate $2N$ normal deviates $\{x_n,y_n\}_{n=1}^N$ and set $c_n=(x_n+\ii y_n)/\sqrt{\sum_k (x_k^2+y_k^2)}$. 

In the following, we consider the case where $\hat{O}_n=|n\rangle\langle n|$ is an electronic population, for which the time-dependent estimator is simply $O_n(c_t)$. There are multiple ways to construct this estimator.\cite{Runeson2023mash} If we are interested in an adiabatic population, then the state projector $\Theta_a(c_t)$ is a natural choice that is consistent with the adiabatic surface the nuclei are evolving on. However, the state projector is not a good estimator for diabatic populations because it does not transform correctly under unitary basis transformations.\cite{Runeson2023mash} 

Another estimator that would transform correctly between bases is the Ehrenfest population $|c_n|^2$, but inserting this choice into Eq.~\eqref{eq:Ot} leads to the wrong long-time equilibrium populations. 
In Ref.~\onlinecite{Runeson2023mash} it was shown that a simple estimator that fulfils both criteria (equivariance under unitary basis transformations and consistency with the quantum-classical equilibrium populations) is
\begin{equation} \label{eq:Phi}
    O_n(c) = \alpha_N |c_n|^2 + \beta_N
\end{equation}
where 
\begin{equation}\label{eq:alphabeta}
\alpha_N = \frac{N-1}{\sum_{k=1}^N(1/k)-1} \qquad \beta_N =\frac{1-\alpha_N}{N}  
\end{equation}
are two scalars that require no more information than the number of states. This is the population estimator that is used in multi-state MASH calculations.\cite{Runeson2023mash}

\subsection{Initial conditions}
What remains to be defined is the choice of initial electronic distribution $\rho_i(c)$ in Eq.~\eqref{eq:Ot}. This distribution is not unique, in the sense that many (quasi)probability distributions will fulfil the initial condition
\begin{equation}\label{eq:target}
    \frac{\int_{|c|=1} \rd c\, \rho_i(c) O_j(c)}{\int_{|c|=1} \rd c\, \rho_i(c)} = \delta_{ij}.
\end{equation}
In the following we consider a few options.

\begin{figure*}
    \includegraphics{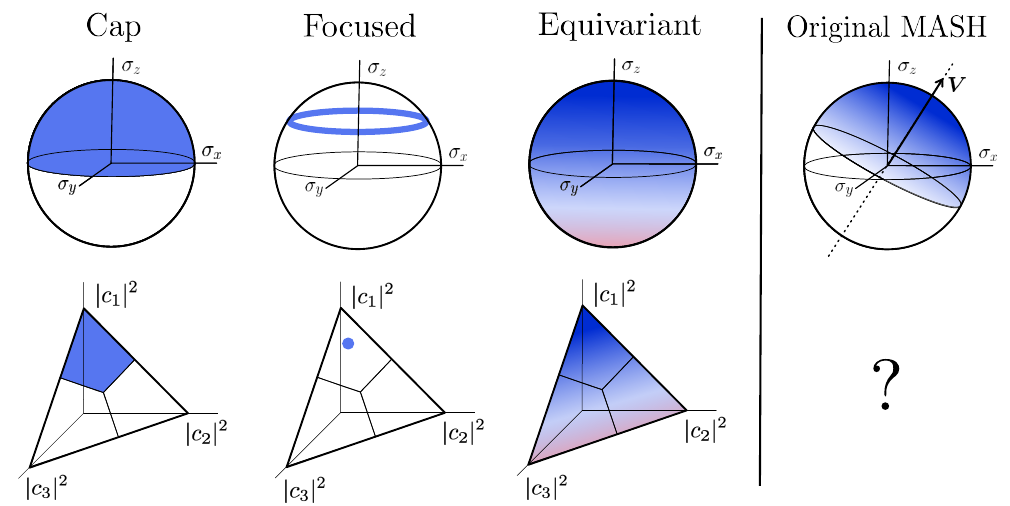}
    \caption{Schematical overview of the MASH initial conditions considered in this article for the case of two (top row) and three (bottom row) states. In the two-state case, $\sigma_x=2\Re c_1^*c_2$, $\sigma_y=2\Im c_1^* c_2$, and the vertical axis $\sigma_z=|c_1|^2-|c_2|^2$ corresponds to polarization in a given diabatic basis, while the tilted axis in the right column corresponds to polarization in the adiabatic basis. Dark blue shading indicates a higher weight and red shading a negative weight. The initial conditions in the first three columns are used together with the equivariant time-dependent observable in Eq.~\eqref{eq:Phi}. The original MASH method uses an alternative prescription that is (so far) limited to two states.}\label{fig:initial}
\end{figure*}

\subsubsection{Cap initial condition}
A simple choice is 
\begin{equation} \label{eq:rhoTheta}
    \rho_i(c)=\Theta_i(c),
\end{equation}
which was used in Ref.~\onlinecite{Runeson2023mash}. What this equation implies is that the initial state is chosen randomly from the region where $|c_i|^2$ is the largest population. The left column in Fig.~\ref{fig:initial} visualizes this region for two and three-level systems. We will refer to these regions as `caps' on the sphere/simplex and consequently to Eq.~\eqref{eq:rhoTheta} as the `cap' initial distribution.

Although the cap distribution has been found to be accurate in a variety of benchmark calculations,\cite{Runeson2023mash} it is not basis-equivariant, unlike the population estimator in Eq.~\eqref{eq:Phi}. A related issue is that the diabatic basis is not unique, which makes the diabatic state projector ambigously defined. As a simple example, consider a three-state system in which we want to start from state 1. Suppose we sample the vector $c=\frac{1}{\sqrt{3}}(\sqrt{1+2\epsilon},\sqrt{1-\epsilon},\sqrt{1-\epsilon})$, where $\epsilon$ is some number in the range $0<\epsilon<1$. This vector has $\Theta_1(c)=1$ and would therefore contribute to the dynamics. But if we define a new set of diabatic basis vectors $|\tilde{1}\rangle=|1\rangle$, $|\tilde{2}\rangle = (|2\rangle + |3\rangle)/\sqrt{2}$, $|\tilde{3}\rangle = (|2\rangle-|3\rangle)/\sqrt{2}$, then for $\epsilon<\frac{1}{4}$ the maximally populated state is $|\tilde{2}\rangle$. Thus, even though the ket of the initial state is unchanged, the same vector $c$ would no longer contribute to the dynamics.

\subsubsection{Focused initial condition}
Perhaps the most intuitive choice for the initial wavefunction corresponding to $|i\rangle\langle i|$ would be to set $c_i=1$ and $c_{j\neq i}=0$, corresponding to a pole on the sphere or a corner of the simplex. This approach is standard in Ehrenfest dynamics and (provided $i$ is an adiabat) in FSSH. However, for MASH it would violate the constraint in Eq.~\eqref{eq:target}. The reason is that in the initial corner of the simplex, $O_i(c)>1$ and $O_{j\neq i}(c)<0$. 

The analogous initial condition for MASH with the correct initial value would be to start from wavefunctions $c$ for which $O_i(c)=1$ and $O_{j\neq i}(c)=0$. Such wavefunctions are confined to circles on the sphere and isolated points on the simplex, as shown in the second column of Fig.~\ref{fig:initial}. The circle (point) is defined by $|c_i|^2 = (1-\beta_N)/\alpha_N$ and $|c_{j\neq i}|^2 = -\beta_N/\alpha_N$. These conditions fix the magnitudes of all components of $c$, leaving the phases to be sampled uniformly from $[0,2\pi)$. The resulting `focused' initial distribution can be written as
\begin{equation}\label{eq:focused}
    \rho_i(c) = \delta\left(|c_i|^2-\frac{1-\beta_N}{\alpha_N}\right)\prod_{j\neq i}\delta\left(|c_j|^2+\frac{\beta_N}{\alpha_N}\right).
\end{equation}

An advantage of this choice is that each trajectory is initialized with physical population observables, so it may be possible to use fewer trajectories than with the cap initial condition to reach statistical convergence. Nevertheless, the focused distribution is also not basis-equivariant. 

\subsubsection{Equivariant initial condition}
In this section, we derive a quasiprobability distribution $\rho_i$ that transforms correctly under unitary basis transformations. To satisfy the condition in Eq.~\eqref{eq:target}, the simplest approach is to try the same functional form as the time-dependent observable in Eq.~\eqref{eq:Phi}, i.e.
\begin{equation}
    \rho_i(c) = a_N |c_i|^2 + b_N
\end{equation}
with some constants $a_N,b_N$ that need not be the same as $\alpha_N,\beta_N$. The resulting $\rho_i(c)$ need not be positive definite since we can multiply each $c$ sampled from the $|c|=1$ sphere by a weight that is positive or  negative.

To evaluate the integral in Eq.~\eqref{eq:target}, we make use of the following moments:
\begin{equation}\label{eq:moments}
    \langle |c_n|^2\rangle = \frac{1}{N}, \qquad
    \langle |c_n|^2|c_m|^2\rangle =\frac{\delta_{nm}+1}{N(N+1)}
\end{equation}
where 
\begin{equation}
    \langle f \rangle \equiv \frac{\int_{|c|=1} \rd c\, f}{\int_{|c|=1} \rd c}.
\end{equation}
These expectation values can be derived using standard formulas for integrals over a sphere.\cite{folland2001}
With the help of the moments in Eq.~\eqref{eq:moments}, Eq.~\eqref{eq:target} reduces to two equations in two unknowns,
with the solution
\begin{equation}
a_N = \frac{N+1}{\alpha_N}, \qquad b_N = -1-a_N\beta_N.
\end{equation}
Note that there is some similarity between the Roman constants and the Greek ones in Eq.~\eqref{eq:alphabeta}. For example, $\alpha_N$ and $\beta_N$ are related through $N\beta_N = 1-\alpha_N$, and likewise $a_N$ and $b_N$ are related through $N b_N = 1-a_N$.
This means that not only $O_n(c)$ but also $\rho_i(c)$ involves a scaling relative to the centre of the simplex: after inserting $\beta_N$ and $b_N$ we get
\begin{equation}
O_n(c) = \frac{1}{N} + \alpha_N\left(|c_n|^2 - \frac{1}{N}\right)
\end{equation}
and
\begin{equation}
\rho_i(c) = \frac{1}{N} + a_N\left(|c_i|^2 - \frac{1}{N}\right).
\end{equation}
The special status of the centre of the simplex is analogous to the special role of the identity operator in phase-space mapping methods.\cite{saller2019jcp,gao2020nonadiabatic}

\subsubsection{Original MASH}
For two-level systems, the original MASH method of Mannouch and Richardson\cite{Mannouch2023mash} uses an alternative prescription in which diabatic observables are first converted to the adiabatic basis, where populations and coherences are then measured with different estimators. Explicitly, for $a\neq a'$ and $b\neq b'$, their prescription is (in our notation)
\begin{align}
    \Tr[|a\rangle\langle a|(0)|b\rangle\langle b|(t)] &\mapsto  \frac{2\int \rd c \, W_a(c) \Theta_a(c)  \Theta_b(c_t)}{\int \rd c} \\
    \Tr[|a\rangle\langle a|(0)|b\rangle\langle b'|(t)] &\mapsto \frac{4\int \rd c  \, \Theta_a(c) c_{b,t}^* c_{b',t}}{\int \rd c} \\
    \Tr[|a\rangle\langle a'|(0)|b\rangle\langle b'|(t)] &\mapsto \frac{6\int \rd c \, c_a^* c_{a'}\, c_{b,t}^* c_{b',t}}{\int \rd c}
\end{align}
where $W_a(c) = 4|c_a|^2-2$ is a weight that goes to zero when the two adiabats have equal populations.
Note that this approach differs from the others above not only in the initial distribution but also in the construction of the time-dependent observable.

\section{Results}

\begin{figure*}
    \centering
    \includegraphics{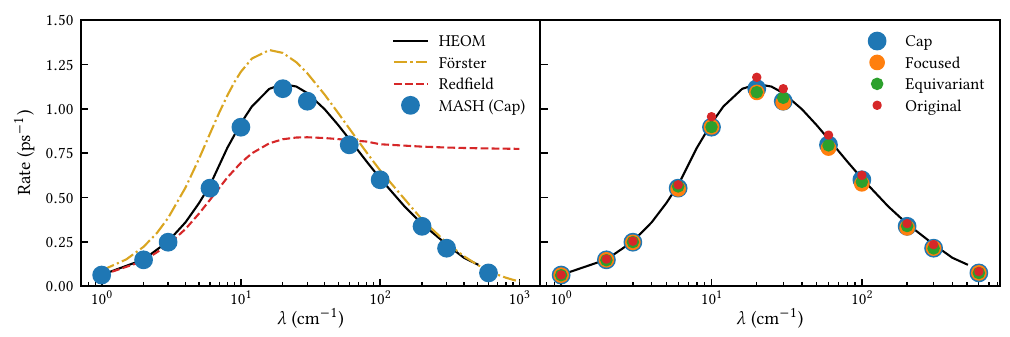}
    \caption{Rate of intersite population transfer in a Frenkel-exciton dimer as a function of the bath reorganization energy. MASH agrees closely with the HEOM benchmark across the entire parameter range (left panel), regardless of the particular choice of initial condition (right panel).}
    \label{fig:dimer}
\end{figure*}

\subsection{Exciton dimer}
To investigate the transition from Redfield to F\"{o}rster-like transfer, we consider a two-site exciton model with $\epsilon_1-\epsilon_2=\SI{100}{cm^{-1}}$ and $J_{12}=\SI{20}{cm^{-1}}$. Each site is coupled to a bath at $T=\SI{300}{K}$ with the Debye spectral density 
\begin{equation}
    J(\omega) = 2\lambda \frac{\omega\omega_{\rm c}}{\omega^2+\omega^2_{\rm c}}
\end{equation}
where $\omega_{\rm c}=\SI{53}{cm^{-1}}$. The quantity of interest is the forward intersite rate $k_{1\to 2}$ as a function of $\lambda$. This model is well-studied in the literature\cite{ishizaki2009unified,ishizaki2009redfield,huo2015electronic,runesonPhD} and therefore allows comparison with a wide range of methods. Quantum mechanical (HEOM) benchmark results have been computed by Ishizaki and Fleming,\cite{ishizaki2009unified} who observed that Redfield theory is accurate for small $\lambda$ but qualitatively wrong for large $\lambda$, whereas F\"{o}rster theory is only valid for large $\lambda$. In their calculations, the forward and backward rate constants were obtained by fitting the population dynamics to the kinetic model 
\begin{align}\label{eq:two-parameter}
    \frac{d}{dt}\langle P_1(t)\rangle &= -k_{1\to 2}\langle P_1(t)\rangle + k_{2\to 1}\langle P_2(t)\rangle \nonumber\\   
    \frac{d}{dt}\langle P_2(t)\rangle &= \phantom{+}k_{1\to 2}\langle P_1(t)\rangle - k_{2\to 1}\langle P_2(t)\rangle 
\end{align}
where the site density matrix was initialized as $|1\rangle\langle 1|$. No secular approximation was applied in the Redfield theory.

To assess the performance of MASH, we have calculated the population dynamics using all three of the initial conditions considered in Section~\ref{sec:mash}, and performed additional calculations with the original version of MASH for comparison. The bath was discretized into 100 modes per site using a standard discretization scheme~\cite{Hele2013masters} and the nuclei were initialized from the classical Boltzmann distribution of an uncoupled bath. The dynamics was averaged over $10^5$ trajectories. 

When extracting the rate, we observed that fitting the population difference $\langle\sigma_z(t)\rangle=\langle P_1(t)-P_2(t)\rangle$ to a single exponential,
\begin{equation}\label{eq:one-parameter}
    \langle \sigma_z(t) \rangle = (\langle \sigma_z(0)\rangle-\langle\sigma_z\rangle_{\rm eq}) \e^{-k_{\rm tot} t} + \langle \sigma_z\rangle_{\rm eq},
\end{equation}
with $k_{\rm tot}=k_{1\to 2}+k_{2\to 1}$ as a fitting parameter, was more stable than using the two-parameter fit in Eqs.~\eqref{eq:two-parameter}. 
Since the nuclear statistics is essentially classical ($k_{\rm B}T<\omega_{\rm c}$), MASH is guaranteed to recover the correct equilibrium value $\langle \sigma_z \rangle_{\rm eq} = \langle P_1 - P_2\rangle_{\rm eq}$, as do HEOM, Redfield and F\"{o}rster theory. So there is no need for an additional free parameter. Once $k_{\rm tot}$ has been extracted from Eq.~\eqref{eq:one-parameter}, the forward rate constant can be calculated as
\begin{equation}
    k_{1\to 2} = k_{\rm tot}\langle P_2\rangle_{\rm eq} = k_{\rm tot}\frac{1}{2}\left(1-\langle\sigma_z\rangle_{\rm eq}\right),
\end{equation}
which follows from the detailed balance relation $k_{1\to 2}/k_{2\to 1}=\langle P_2\rangle_{\rm eq}/\langle P_1\rangle_{\rm eq}$.
To ensure a fair comparison, we have also recalculated the Redfield and HEOM population dynamics (using the Pyrho open source software package\cite{pyrho}), and applied the same fitting procedure to those. This was found to lead to slighly ($<10$\%) different rates compared to Ref.~\onlinecite{ishizaki2009unified}.

Figure~\ref{fig:dimer} shows our results together with the F\"{o}rster theory rates from Ref.~\onlinecite{ishizaki2009unified}.
We find that MASH agrees closely with HEOM across the entire parameter range (left panel), including the Redfield and F\"{o}rster-type regimes.
Moreover, all four versions of MASH lead to essentially the same rates (right panel). This is interesting because
the cap initial condition has previously been found to be less accurate than the original version of MASH for a spin-boson model in the Marcus inverted regime.\cite{Runeson2023mash}  In the present calculations, the region $2\lambda<\epsilon_1-\epsilon_2=\SI{100}{cm^{-1}}$ is formally in the inverted regime, but the different initial conditions nevertheless lead to similar behaviour. We reach the same conclusion when we convert the Frenkel-exciton model into a spin-boson model with matching bias and total reorganization energy, which we find does not noticeably change either the HEOM or the MASH rates. Hence, all four versions of MASH can be regarded as reliable in the present inverted regime. The inverted regime considered in Ref.~\onlinecite{Runeson2023mash} was more challenging owing to its larger bias (20 times the diabatic coupling matrix element rather than the 5 times considered here), and in that regime the original version of MASH is to be preferred.


\begin{figure*}
    \includegraphics{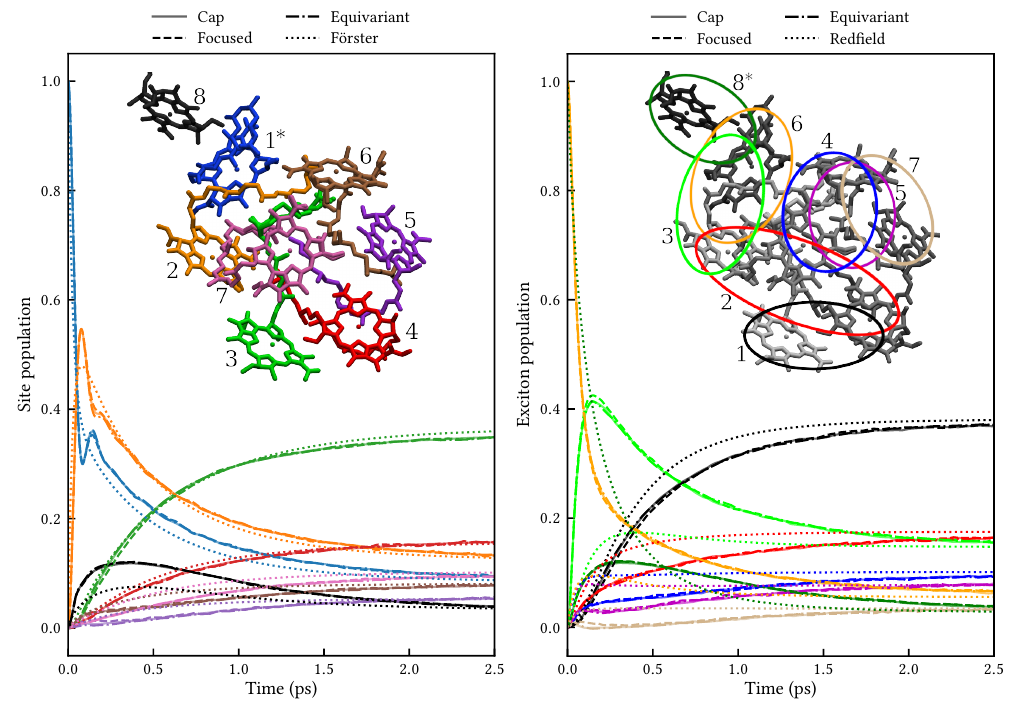}
    \caption{Population dynamics in FMO at $\SI{300}{K}$ comparing MASH with different initial conditions to well-established rate theories. Left: dynamics in the site basis after an initial excitation of site 1. The three MASH initial conditions lead to identical results and agree qualitatively with F\"{o}rster theory at long times. The inset shows the site labels using the same colouring as for the data curves. Right: dynamics in the exciton basis after an initial excitation of exciton 8. The three MASH initial conditions lead to similar results and predict notably slower transfer than (secular) Redfield theory. The inset depicts qualitatively the spatial extent of the exciton states and their labels in order of increasing energy.}\label{fig:fmo}
\end{figure*}

\subsection{Eight-site Fenna--Matthews--Olson complex}
Another well-known benchmark system for exciton energy transfer is the Fenna--Matthews--Olson complex found in green sulfur bacteria. We have previously demonstrated that MASH (with cap initial conditions) agrees closely with HEOM for a standard seven-site FMO model with a Debye spectral density.\cite{Runeson2023mash} Here, we consider a more challenging (and realistic) eight-site model with a structured spectral density extracted from fluorescence line narrowing experiments.\cite{wendling2000electron} The resulting bath has a reorganization energy of $\lambda=\SI{45}{cm^{-1}}$. The intersite couplings and average site energies (shown in Table~\ref{tab:HS}) were obtained from electrostatic calculations for the FMO complex of \emph{Prosthecochloris aestuarii}.\cite{amBusch2011fmo} Static disorder was included by sampling the site energies with the Gaussian widths shown in Table~\ref{tab:FWHM}, which were calculated by M\"{u}h \emph{et al}.\cite{mueh2007pnas}

We are not aware of any fully quantum benchmarks for this model, so instead we compare MASH to F\"{o}rster and Redfield theory. These methods have a long history in modelling the dynamics of FMO.\cite{Milder2010,cao2020review,renger2021detailedbalance} 
Here, we calculate the dynamics in the site basis using F\"{o}rster theory and the dynamics in the exciton basis using Redfield theory within the secular approximation. In each basis, the dynamics is therefore simply a propagation of the populations with a constant rate matrix. For simplicity, we start from an excitation localized on a single site (for the site basis calculation) or on a single exciton (for the exciton basis calculation). The F\"{o}rster and Redfield dynamics were averaged over 1000 samples of the site energies to account for static disorder.
In the MASH calculations, the bath was discretized into 100 modes per site using an equally spaced grid up to $\omega_{\rm max}=\SI{500}{cm^{-1}}$, and the modes were initialized from the classical Boltzmann distribution of an uncoupled bath at $\SI{300}{K}$. The dynamics were averaged over $10^6$ trajectories for the cap and equivariant initial conditions and $10^5$ for the focused initial condition to ensure tight convergence.

The left panel of Fig.~\ref{fig:fmo} shows the site populations after an initial excitation of site 1. All three MASH initial conditions give indistinguishable results. Apart from a transient ($<\SI{0.5}{ps}$) coherence between sites 1 and 2, the dynamics is essentially rate-like. Although F\"{o}rster theory does not capture the coherence and differs from MASH at short times, in particular for site 8, it agrees qualitatively with MASH at longer times. This observation is consistent with previous studies for simpler FMO models,\cite{Wu2012forster,Wilkins2015masters} where F\"{o}rster theory was found to be qualitatively reliable in comparison with exact benchmark results, despite several site couplings being as strong as \SI{90}{cm^{-1}}. The reason is likely that the strong couplings only matter for the first $\sim\SI{100}{fs}$, whereas on the $\sim\SI{1}{ps}$ timescale the population transfer is controlled by the weaker couplings for which F\"{o}rster theory is accurate.

The right panel of Fig.~\ref{fig:fmo} shows the exciton populations after an initial excitation of exciton 8. This exciton state is spatially located on sites 8 and 1, and has been identified as one of the dominant pathways when captured photon energy enters the FMO complex from the baseplate of the chlorosomes.\cite{cao2020review} 
Again, all MASH initial conditions lead to similar dynamics up to a slight difference that washes out within \SI{1}{ps}. Notably, the overall transfer is significantly slower than in Redfield theory, by roughly a factor of 2. This observation is consistent with a previous study using a phase-space mapping of the electronic states,\cite{runeson2022fmo} where it was shown that even though the Markovian approximation is valid for the present bath, the system--bath coupling is too large for second-order perturbative approaches like Redfield theory to be reliable (see Figs.~S2 and S5 of Ref.~\onlinecite{runeson2022fmo}). 
Since MASH has the additional advantage of relaxing to the correct long-time limit, we expect it to be more accurate than those previous mapping calculations. Note, however, that MASH can experience negative populations for intermediate times. In the present calculations, exciton state 7 becomes slightly negative between 0.1 and 0.3 ps with the `cap' and `equivariant' initial conditions. This could be a real effect or due to insufficient sampling. Currently, neither of the versions of MASH guarantees complete positivity of the system density matrix except in the long-time limit. (For two states, the original MASH gives strictly non-negative populations only in the adiabatic basis.) The quantum-jump correction\cite{Mannouch2023mash,Lawrence2023mash} may help to alleviate this deficiency in future work.

\begin{table}
\centering
\caption{Average site energies and couplings for FMO\cite{amBusch2011fmo} in units of ${\rm cm}^{-1}$.\\}
\label{tab:HS}
\begin{tabular}{c|d{3.1}d{3.1}d{3.1}d{3.1}d{3.1}d{3.1}d{3.1}d{3.1}}
 Site & \multicolumn{1}{c}{1}     & \multicolumn{1}{c}{2}    & \multicolumn{1}{c}{3}     & \multicolumn{1}{c}{4}     & \multicolumn{1}{c}{5}    & \multicolumn{1}{c}{6}     & \multicolumn{1}{c}{7}     & \multicolumn{1}{c}{8}      \\ 
\hline
1 & \multicolumn{1}{c}{310} & -94.8 & 5.5   & -5.9  & 7.1   & -15.1 & -12.2 & 39.5   \\
2 &       & \multicolumn{1}{c}{230} & 29.8  & 7.6   & 1.6   & 13.1  & 5.7   & 7.9    \\
3 &       &       & \multicolumn{1}{c}{0} & -58.9 & -1.2  & -9.3  & 3.4   & 1.4    \\
4 &       &       &       & \multicolumn{1}{c}{180} & -64.1 & -17.4 & -62.3 & -1.6   \\
5 &       &       &       &       & \multicolumn{1}{c}{405} & 89.5  & -4.6  & 4.4    \\
6 &       &       &       &       &       & \multicolumn{1}{c}{320} & 35.1  & -9.1   \\
7 &       &       &       &       &       &       & \multicolumn{1}{c}{270} & -11.1  \\
8 &       &       &       &       &       &       &       & \multicolumn{1}{c}{505}
\end{tabular}
\end{table}

\begin{table}
    \centering
    \caption{Gaussian widths (full width at half maximum) of the site energies\cite{mueh2007pnas} in ${\rm cm}^{-1}$.\\}
    \label{tab:FWHM}
    \begin{tabular}{c|cccccccc}
        Site & 1 & 2 & 3 & 4 & 5 & 6 & 7 & 8  \\ \hline
        FWHM & 60 & 100 & 60 & 60 & 120 & 120 & 120 & 100
    \end{tabular}
\end{table}

\section{Conclusions}

In this article, we have shown by comparison with exact results that MASH correctly captures the transition from the Redfield to F\"{o}rster regimes for an exciton dimer. This is the case no matter if one uses the original two-state observables or the equivariant estimators in multi-state MASH. In conjuction with the recent finding that MASH recovers Marcus theory in the diabatic limit,\cite{Lawrence2023mash} our results further establish MASH as a generally reliable rate theory across several relevant parameter regimes. Since it additionally relaxes to the correct equilibrium populations for excitonic systems in classical environments, in contrast to any other nonadiabatic dynamics method we are aware of, and since it is applicable to systems described by general anharmonic interaction potentials, we would argue that MASH is a practical tool that is capable of capturing almost all of the relevant ingredients of exciton transfer. (It has yet to be generalized to include quantum mechanical effects in the nuclear motion, which is a work in progress.)

For a challenging model of FMO including static disorder and an experimental spectral density, we find that MASH agrees qualitatively with F\"{o}rster theory (apart from a short transient coherence in the site basis), even though several inter-site couplings are expected to be beyond the range of applicability of the golden rule. In the exciton basis, MASH differs from Redfield theory in its slower energy transfer timescale, confirming findings from spin-mapping methods\cite{runeson2020,runeson2022chimia} that the system--bath coupling is too large to treat as a perturbation.\cite{runeson2022fmo}

We have also described and resolved an important issue regarding the initial conditions in multi-state MASH. For the present systems, we find that the results are virtually identical for various different choices of the initial conditions. Although the situation would likely be different for applications in excited-state photochemistry, we conclude that for the condensed-phase environments considered here one may use whichever initial condition is more practical. A previous calculation for a spin-boson model in the Marcus inverted regime has found that the `cap' initial condition in multi-state MASH and the original MASH method give different relaxation timescales,\cite{Runeson2023mash} but for the present dimer model there is no noticeable difference between the two methods even for model parameters that correspond to the inverted regime.

\section*{Conflicts of interest}
There are no conflicts to declare.

\section*{Acknowledgements}
J.E.R. was funded by a mobility fellowship from the Swiss National Science Foundation. T.P.F. was supported by the U.S. Department of Energy, Office of Science, Basic Energy Sciences, CPIMS Program Early Career Research Program under Award DE-FOA0002019.

\bibliography{runerefs}

\end{document}